\documentclass[11pt]{article}
\usepackage{graphicx}
\usepackage{color}
\begin{document}

\title{
Production of hot nuclei in Fermi energy domain: from peripheral 
to central collisions.}

\author{ 
M. Veselsky
\thanks{Phone: (979)-845-1411, fax: (979)-845-1899, e-mail: veselsky@comp.tamu.edu}
\thanks{On leave of absence from Institute of Physics of SASc, Bratislava, 
Slovakia}
\\
{\small Cyclotron Institute, Texas A\&M University, 
College Station, TX 77843.} 
}

\maketitle

\begin{abstract}
{
The production mechanism of highly excited nuclei in the Fermi energy domain 
is investigated. 
A pheno\-menological approach, based on the exciton model, is used for the 
description of pre-equilibrium emission. A model of deep inelastic transfer is 
employed for the peripheral collisions in the post-pre-equilibrium stage. 
An approach to describe more central collisions is proposed. 
A geometric overlap formula is employed in a way suitable for given 
energy domain. A simple geometric approach describing the interaction of 
participant and spectator zones is used to determine the incomplete 
fusion channel. Excitation energies of both fragments are determined. 
Results of the calculation are compared to available experimental data 
and an overall satisfactorily agreement is obtained. 
The model's ability to describe the production of the hot nuclei 
can be employed in the study of multifragmentation 
and/or in the production of rare beams. 
}
\end{abstract}

A detailed knowledge of the production mechanism of hot nuclei is desirable 
for the understanding of multifragmentation. The peripheral collisions of 
heavy ions in the Fermi energy domain demonstrate that the nucleon exchange 
is a dominating production mechanism of the highly excited quasiprojectiles. 
In more violent collisions, the processes leading to the mid-velocity emission 
start to play an important role. The influence of the pre-equilibrium emission 
and fragmentation-like processes becomes an important issue. There exists 
a considerable amount of experimental data from the reactions of heavy ions 
in the Fermi energy domain ( for review see e.g. refs. \cite{ExpRev,LGPT} ). 
Various transport and molecular dynamics models are generally used for 
a theoretical description ( for review see e.g. ref. \cite{TrRev} and 
references therein ). In this article, a framework based on simple 
phenomenological assumptions will be presented and its ability to describe 
the experimental data will be demonstrated. 

The pre-equilibrium emission ( PE ) is a process where fast particles are 
emitted prior to the equilibration of the system. Emission of pre-equilibrium 
particles in reactions induced by nucleons and light particles was 
theoretically explained using the exciton model \cite{Griff} or using the master 
equations \cite{HMB}. For reactions induced by heavy ion beams 
a model of nucleon exchange was developed \cite{RanVan}. In the present work, 
we use a phenomenological description \cite{VeZPA} based on similar assumptions 
as the exciton model. The probability of pre-equilibrium emission 
is evaluated using the formula 
\begin{equation}
\label{ppre}
P_{pre}(n/n_{eq})=1-e^{-\frac{(n/n_{eq}-1)^2}{2\sigma ^2}}
\end{equation}
for $n\le n_{eq}$ and is assumed zero for $n>n_{eq}$, 
where $n$ is the number of excitons at given stage and $n_{eq}$ is the 
the number of excitons in the equilibrium configuration for given 
excitation energy. The $\sigma$ is a free parameter. The basic assumption 
leading to eq. ( \ref{ppre} ) is the dependence of P$_{pre}$ on the 
ratio $n/n_{eq}$ as suggested in ref. \cite{Boh}. An initial exciton 
number is equal to the mass number of the beam. The equilibrium 
number of excitons is calculated according to the formula \cite{Boh} 
\begin{equation}
\label{neq}
n_{eq}=2\hbox{ }g\hbox{ }T\hbox{ }\ln2
\end{equation}
where $g$ is the one particle level density at the Fermi energy and $T$ is the 
nuclear temperature determined as $T^2=U/\tilde{a}$, where $\tilde{a}$ is the 
level density parameter and $U$ is the excitation energy. At every 
emission step, a random number between zero and one is generated. If the random 
number is smaller than P$_{pre}$ a pre-equilibrium particle is emitted. 
The emission of neutron, proton and $\alpha$-particle is considered 
and the Weisskopf-Ewing emission widths are used. The Maxwellian 
spectrum of kinetic energy with the apparent temperature \cite{FuMo} 
\begin{equation}
\label{tappa}
T_{app}= [\frac{2.5}{A_{P}}(E_{P}-V_{C})]^{1/2}
\end{equation}
is assumed, where A$_P$ is the projectile mass number, E$_P$ is the projectile 
energy and V$_C$ is the Coulomb barrier. The emission angle is determined 
according to the formula \cite{Park}
\begin{equation}
\label{thdprk}
\frac{d\sigma} {d\Omega}=K \hspace{2mm} exp(\frac {-\theta}{\Delta \theta})
\end{equation}
where $\Delta \theta = \frac {2\pi} {kR_{CN}}$ , $R_{CN}$ is the radius 
of the compound nucleus and $k$ is the wave number of the emitted particle. 
After emission, the exciton number is increased by a value obtained using 
the formula 
\begin{equation}
\label{dltn}
\Delta n=A_{pre}\hbox{ }\frac{\kappa}{\beta_{rad}}
\end{equation}
where A$_{pre}$ is the mass of emitted particle, $\beta_{rad}$ is the 
radial velocity in the contact configuration at a given angular 
momentum and $\kappa$ is a free parameter. 
If no pre-equilibrium emission occurs at a given emission stage, 
the pre-equilibrium stage is finished. 

As a next step, the interaction of the projectile and target is 
considered. As shown in several works \cite{TaGo,VePRC}, the model of deep 
inelastic transfer ( DIT ) describes well the peripheral collisions where the 
relative motion is mostly tangential. With decreasing angular momentum 
the radial motion becomes more intense and violent scenarios 
should be considered. At low energies several MeV/nucleon above the Coulomb 
barrier an incomplete fusion was observed by detecting evaporation 
residues \cite{Park,SlRes,Si-Wil} or fission fragments \cite{Tubbs}. The 
framework of angular momentum windows was introduced by Wilczynski et al. 
\cite{Wilcz}. At very high projectile energies the abrasion-ablation model 
\cite{Gosset} is used widely to describe the fragmentation phenomena. 
In the proposed description, different scenarios are employed depending 
on angular momentum. At large angular momenta, the model of 
deep inelastic transfer is used. In the more central collisions, 
where stationary di-nuclear configuration cannot be created, the 
framework of a geometric overlap is used. In the near-central collisions 
a model of Wilczynski \cite{Wilcz} is employed. 

Since pre-equilibrium emission occurs prior to the fragmentation stage, it is 
necessary to reconstruct the post-pre-equilibrium projectile-target 
configuration. It is assumed, according e.g. to the conclusions of work 
\cite{RanVan} that pre-equilibrium particles are mostly emitted from the 
( usually light ) projectile and propagate through the target. The emitted mass 
and charge are subtracted from the projectile. The excitation energy of the 
target is set equal to the sum of kinetic energies of the emitted 
pre-equilibrium particles, in accordance to the assumption that one 
nucleon-nucleon collision occurs during the propagation of the particle 
through the target.  

For every event the Monte Carlo DIT code of Tassan-Got \cite{TaGo} is used. 
In the case when di-nuclear configuration is created, deep inelastic transfer 
takes place and excited quasi-projectile and quasi-target are 
created. In the cases, where overlap of nuclei is too deep, the more violent 
scenario \hbox{( here} arbitrarily denoted as realistic geometric fragmentation 
- RGF ) is considered. The geomtric overlap formula of the abrasion-ablation 
model \cite{Gosset} is used. It is clearly unrealistic 
to assume that the projectile propagates along straight line determined by the 
asymptotic value of impact parameter. Instead, a minimum distance between 
projectile and target in the Coulomb scattering of two point-like charges 
is used. Thus, one participant and one or two spectator zones are 
created. 

The charges of the spectators are determined according to the 
combinatorial formula \cite{Fried,Gaim} 
{\footnotesize 
\begin{equation}
\label{ffried}
P(Z_{iS}) = (^{Z_{i}}_{Z_{iS}})(^{N_{i}}_{N_{iS}})/(^{A_{i}}_{A_{iS}})
\end{equation}
}
where i=P(T) for the projectile ( target ), A$_{P(T)}$, Z$_{P(T)}$, N$_{P(T)}$ 
are the mass number, charge and neutron number of the projectile ( target ) 
and A$_{P(T)S}$, Z$_{P(T)S}$, N$_{P(T)S}$ are mass, charge and the neutron 
number of the projectile ( target ) spectator. The charge of participant zone 
is determined assuming charge conservation. 

In the Fermi energy domain one can 
assume that the participant zone can be captured by either projectile or target 
spectator zone. In order to decide between these two, the volume occupied 
by the neighboring nucleons within the reach of nuclear interaction ( 1 fm ) 
is determined in both spectators. The volume is approximated by a 1 fm thick slice 
of the sphere. The number of neighboring nucleons ( $A_{NS}$ ) is then 
determined using a normal distribution centered at the value exactly 
corresponding to the volume with the standard deviation equal 
to $\sqrt{A_{NS}}$. The numbers of neighboring nucleons are compared and the 
participant zone is captured by the spectator with more neighboring nucleons. 
Such a procedure reflects the saturation of the nuclear force. 
The capturing spectator and the captured participant zone form a very hot 
fragment. The other spectator is much colder. In this case, the excitation 
energy is determined using the formula 
\begin{equation}
\label{xdiss}
E^{*}_S=x\hbox{ }A_{NS}\hbox{ }(\frac{E_{P'}}{A_{P'}}-V_{C'})\hbox{ }
\frac{<s>}{\lambda}
\end{equation}
where E$_{P'}$ and A$_{P'}$ are the kinetic energy and the mass number of the 
effective projectile after pre-equilibrium emission, V$_{C'}$ is the Coulomb 
barrier, $<s>$ is the mean effective path of the nucleon along the spectator 
trajectory within the slice, $\lambda$ is the mean free path between the 
nucleon-nucleon collisions in the nucleus ( a value 6 fm is chosen ) and 
$x$ is a random number between zero and one. 

The energy and the angle of the spectator is determined using the formula 
of Matsuoka et al. \cite{Matsu} based on the Serber approximation. The formula 
reads 
\begin{equation}
\label{fmatsu}\frac{d^{2}\sigma}{dE_{a}d\Omega_{a}} = \frac {
(E_{a}E_{b})^{1/2} } { (2\mu B_{P'} + 2m_{a}^{2}E_{P'}/m_{P'} + 2m_{a}E_{a} -
4(m_{a}^{3}E_{P'}E_{a}/m_{P'})^{1/2}cos\theta )^{2} }
\end{equation}
where it is the fragment $a$ which flies away and the fragment $b$ which
fuses with the other nucleus, $E_{a}$ and $E_{b}$ are their kinetic energies, 
$ B_{P'} $ is the binding energy of $a$ and $b$ in $P$, $\mu$ is the reduced 
mass of the system $a+b$ , $m_{P'},m_{a},m_{b}$ are the masses of P',a,b and 
$\theta$ is the emission angle of a with respect to the direction of P'.
The kinetic energy and angle of a hot fragment are determined using energy 
and momentum conservation laws. An intrinsic angular momentum 
is calculated using a mean radial distance and momentum of the 
participant zone relative to the capturing spectator in the configuration 
of minimum distance between the projectile and target.  

In order to describe also the inverse kinematics, namely when the projectile 
is heavier than the target, the system is transformed into the inverse frame 
where the projectile becomes a target and vice versa. Then the calculation 
proceeds as described above and the final kinematic properties of the 
reaction products are obtained after a re-transformation into the lab frame. 

\begin{figure}[!htbp]
\centering
\vspace{5mm}
\includegraphics[width=6.cm,height=5.cm]{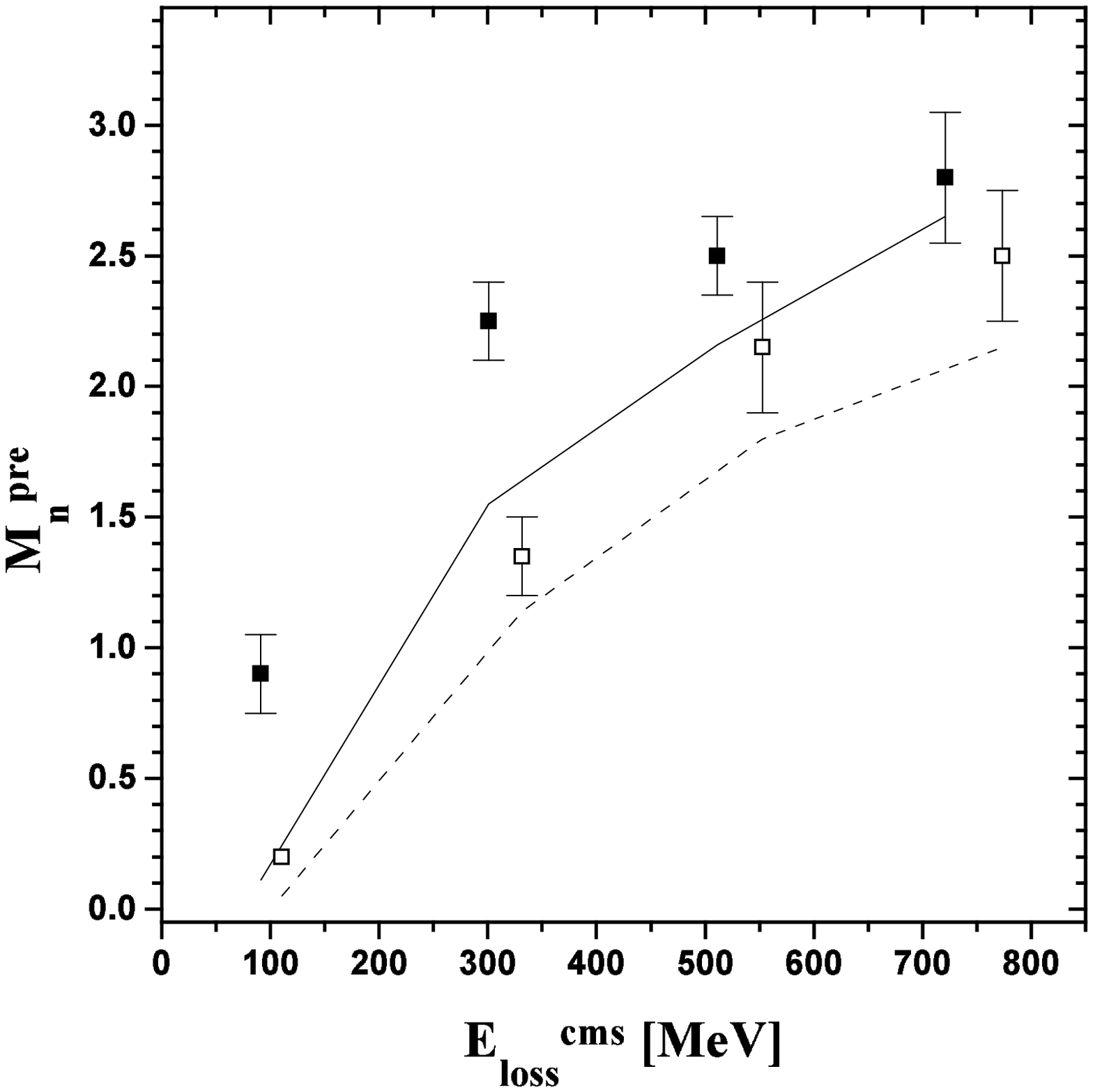}
\caption{ \footnotesize 
Experimental \cite{Agni} ( symbols ) and 
calculated \hbox{( lines )} mean multiplicities 
of pre-equilibrium neutrons as a function of kinetic energy loss 
of the projectile-like fragment. Solid squares - experimental multiplicities 
measured in reaction of 35 MeV/nucleon $^{48}$Ca beam with $^{112}$Sn target, 
open squares - ditto for 35 MeV/nucleon $^{40}$Ca beam, 
solid line -  calculated multiplicities in reaction 
of 35 MeV/nucleon $^{48}$Ca beam with $^{112}$Sn target, 
dashed line - ditto for 35 MeV/nucleon $^{40}$Ca beam. 
}
\label{fgagni}
\end{figure}

The model of pre-equilibrium emission was compared to the results of 
work \cite{Agni} where a multiplicity of the pre-equilibrium particles 
was determined in coincidence with projectile-like 
fragments ( PLFs ) in the reactions of Ca beams with $^{112}$Sn target 
at 35 MeV/nucleon. In Fig. 1 are given the values of pre-equilibrium neutron 
multiplicity in reaction $^{40,48}$Ca+$^{112}$Sn for several bins of  
kinetic energy loss of the projectile-like fragment. The solid ( open ) squares 
represent the results of work \cite{Agni} and the lines represent 
the results of the calculation. The agreement is quite good. The parameters 
$\sigma$=0.25 and $\kappa$=0.3 were used in the calculation. The same values 
of $\sigma$ and $\kappa$ were tested in other reactions and lead to 
results which track well with the results of experimental works 
where multiplicities of pre-equilibrium particles were determined 
in coincidence with heavy residues or fission fragments \cite{Holub,VeZPA} 
or in coincidence with the quasi-projectile \cite{VePRC}. 

\begin{figure}[!htbp]
\centering
\vspace{5mm}
\includegraphics[width=8.cm,height=8.cm]{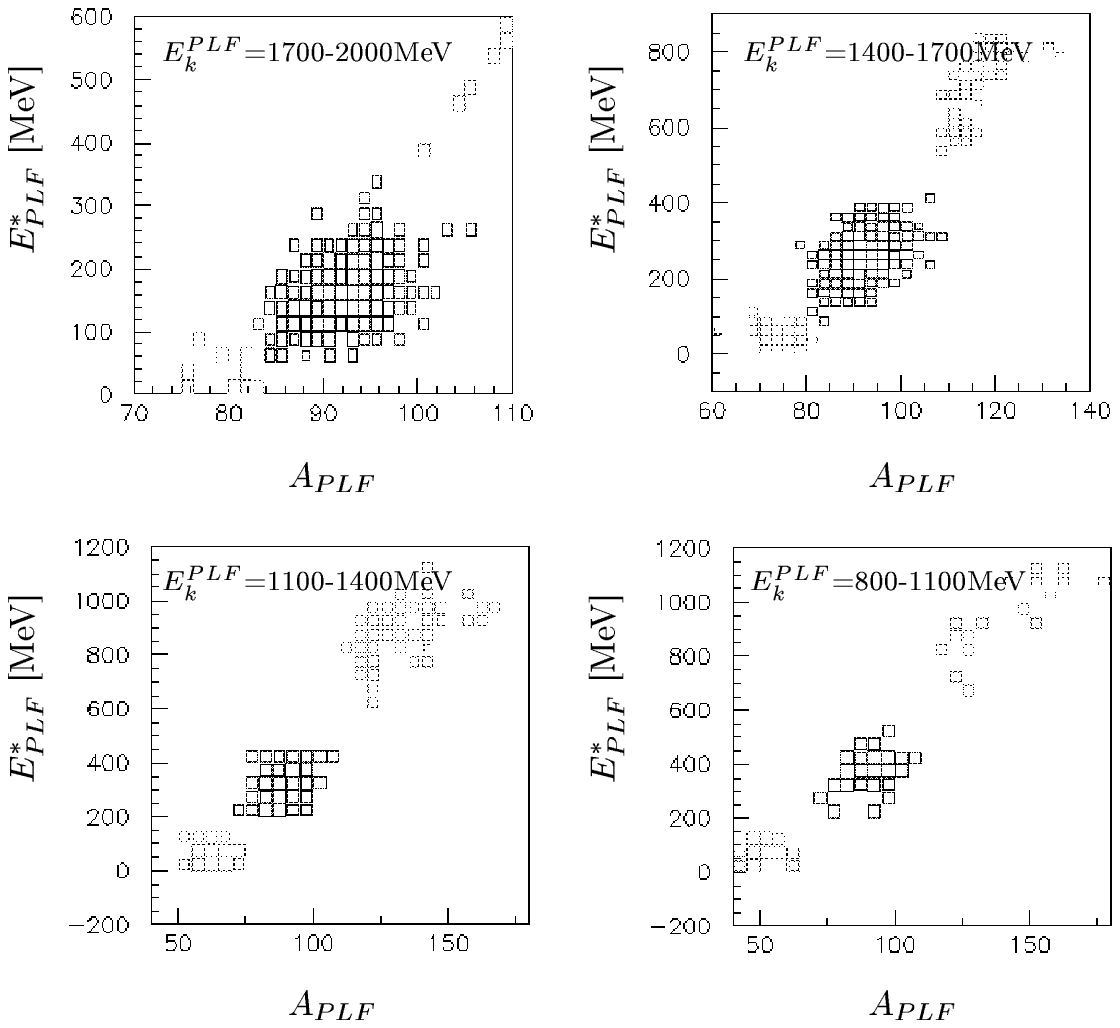}
\caption{ \footnotesize 
Calculated correlation between the primary mass and the excitation energy of 
a projectile-like fragment in the reaction $^{93}$Nb with $^{116}$Sn 
at 25 MeV/nucleon. Four different bins of $E_{k}^{PLF}$ are defined. 
Black squares - PE+DIT events, grey squares - PE+RGF events. 
}
\label{fgcasi}
\end{figure}

In the recent experimental work \cite{Casini} a linear correlation 
between the primary mass of the projectile-like fragment and 
the net mass loss due to the de-excitation was reported in the 
nearly symmetric reactions of $^{93}$Nb with $^{116}$Sn at 25 MeV/nucleon 
in both normal and inverse kinematics for different dissipation bins. 
The net mass loss increases with the primary mass of the projectile-like 
fragment. With increasing dissipation this trend occurs in the still 
broader range of primary masses. Since the net mass loss is possibly correlated 
to the excitation energy of the hot primary projectile-like nucleus, 
one may expect similar trend also for excitation energy. Such a trend 
is a possible signal of the breakdown of the concept of deep inelastic 
transfer. In Fig. 2 is given a calculated correlation between the excitation 
energy and the mass of the hot projectile-like nucleus for different bins of 
kinetic energy. One can see that the calculation follows the experimental 
trend. At masses close to the beam the deep inelastic transfer 
takes place but the range of primary masses is quite narrow. To achieve 
larger mass change a more violent collision should occur. When the target 
strips a part of the projectile, the projectile-like fragment remains 
relatively cold. Hot projectile-like fragments are produced if a part 
of the target is picked-up by the projectile. 

The production of heavy residues was studied recently by Skulski et al. 
\cite{Skul} in the reaction $^{86}$Kr+$^{197}$Au at projectile energy 35 
MeV/nucleon. A spectrum of kinetic energies of the target-like fragments 
( TLFs ) was measured at angles 9 $^{\circ}$ - 46 $^{\circ}$ in coincidence 
with projectile-like fragments with Z $\ge$ 25 
at angles 2 $^{\circ}$ - 8 $^{\circ}$. 
The spectrum exhibits two humps, one of them is the low energy component 
at energies below 40 MeV and the second one is located between 50 and 100 MeV. 
Fig. 3a shows the calculated spectrum of kinetic energies for the 
same reaction using identical angle cuts and charge threshold. 
One can see that similar range of the kinetic energies is covered. 
The double-humped structure is not as clearly pronounced as in the experiment. 
As may be seen in Fig. 3b, the double-humped shape can be explained 
by a correlation between the kinetic energy and the angle. 
The low energy particles are emitted at large angles and an additional energy 
losses or re-scattering can be caused by interaction with the target. 
The high energy part is possibly influenced by decreasing mass and charge and 
increasing excitation energy of the coincident projectile-like fragment what 
leads to the decrease of detection probability because of the charge threshold. 

\begin{figure}[!htbp]
\centering
\vspace{5mm}
\includegraphics[width=8.cm,height=5.cm]{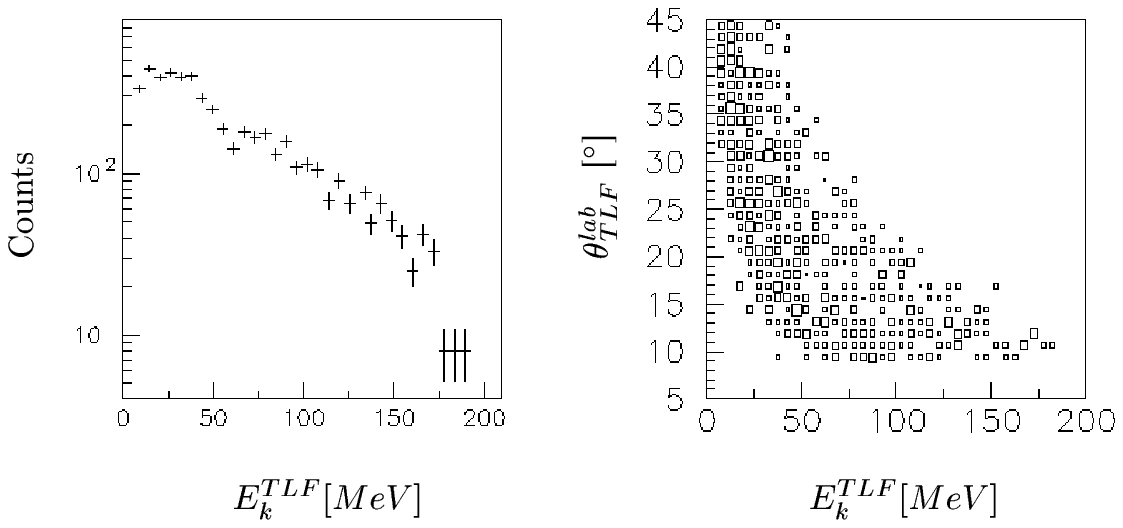}
\caption{ \footnotesize 
Calculated kinetic energy spectrum of target-like fragments 
in reaction $^{86}$Kr+$^{197}$Au at projectile energy 35 
MeV/nucleon and the correlation of $\theta_{TLF}$ and $E_{k}^{TLF}$ 
for the same reaction. For details see text. 
}
\label{fgskul}
\end{figure}

The production of heavy residues in inverse kinematics was measured recently 
by Souliotis et al. \cite{Sou} in the reaction $^{197}$Au+$^{nat}$Ti at 
projectile energy 20 MeV/nucleon. Fig. 4 shows the measured yields 
of heavy residues at the forward angles as a function of A and Z. 
The solid line represents the calculated centroids of Z 
for a given residue mass. The code GEMINI \cite{GEM} was used for the 
de-excitation stage. One can see that the calculation follows the experimental 
trend quite well. In this case the participant zone sticks almost exclusively 
to the heavy projectile and the incomplete fusion leads 
to the production of neutron-deficient residues with masses close to the beam. 

The measurement of the production of intermediate mass fragments ( IMFs )
in symmetric collisions $^{58}$Fe,$^{58}$Ni+$^{58}$Fe,$^{58}$Ni at 30 MeV/u 
\cite{Ramak} determined three different sources of IMFs, the moderately excited 
projectile(target)-like source at velocities near the projectile (target) 
and the highly excited source at velocities near the center of mass velocity. 
Fig. 5 shows a correlation between the excitation energy and the velocity 
of reaction products in the lab frame. Both projectile-like 
and target-like nuclei are included in the plot. One can  
identify three sources analogous to the ones seen in experiment. 
The projectile- and target-like sources are moderately excited. 
The third source with the average velocity close to the mid-velocity 
is highly excited. Both the PLFs and TLFs are included 
in this source. The difference of the isospin of IMFs from different 
sources which was experimentally observed may possibly be explained by 
different excitation energies of the sources. 
As follows from refs. \cite{isodist,mutemp}, the isospin dependences 
of fragment yield ratios are more flat at high excitation energies and 
the sensitivity to isospin of the source is weaker. 

\begin{figure}[!htbp]
\centering
\vspace{5mm}
\includegraphics[width=7.cm,height=5.cm]{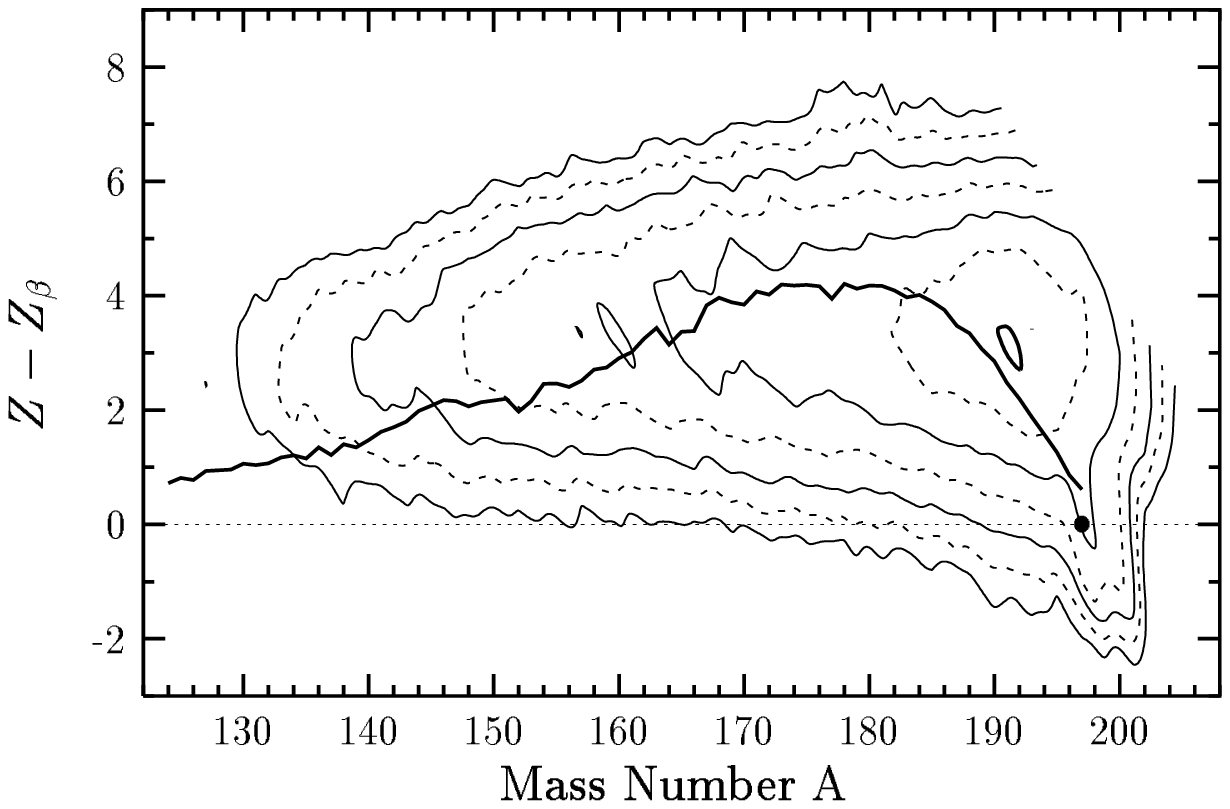}
\caption{ \footnotesize 
Measured yields \cite{Sou} of heavy residues at the forward angles 
in the reaction $^{197}$Au(20 MeV/nucleon)+$^{nat}$Ti as a function of A and Z. 
Z is expressed relative to the line of $\beta$-stability. 
Solid line - calculated centroids of the fragment charge for given residue 
mass ( the code GEMINI \cite{GEM} was used for \hbox{de-excitation ).} 
}
\label{fgsoul}
\end{figure}

\begin{figure}[!htbp]
\centering
\vspace{5mm}
\includegraphics[width=5.cm,height=5.cm]{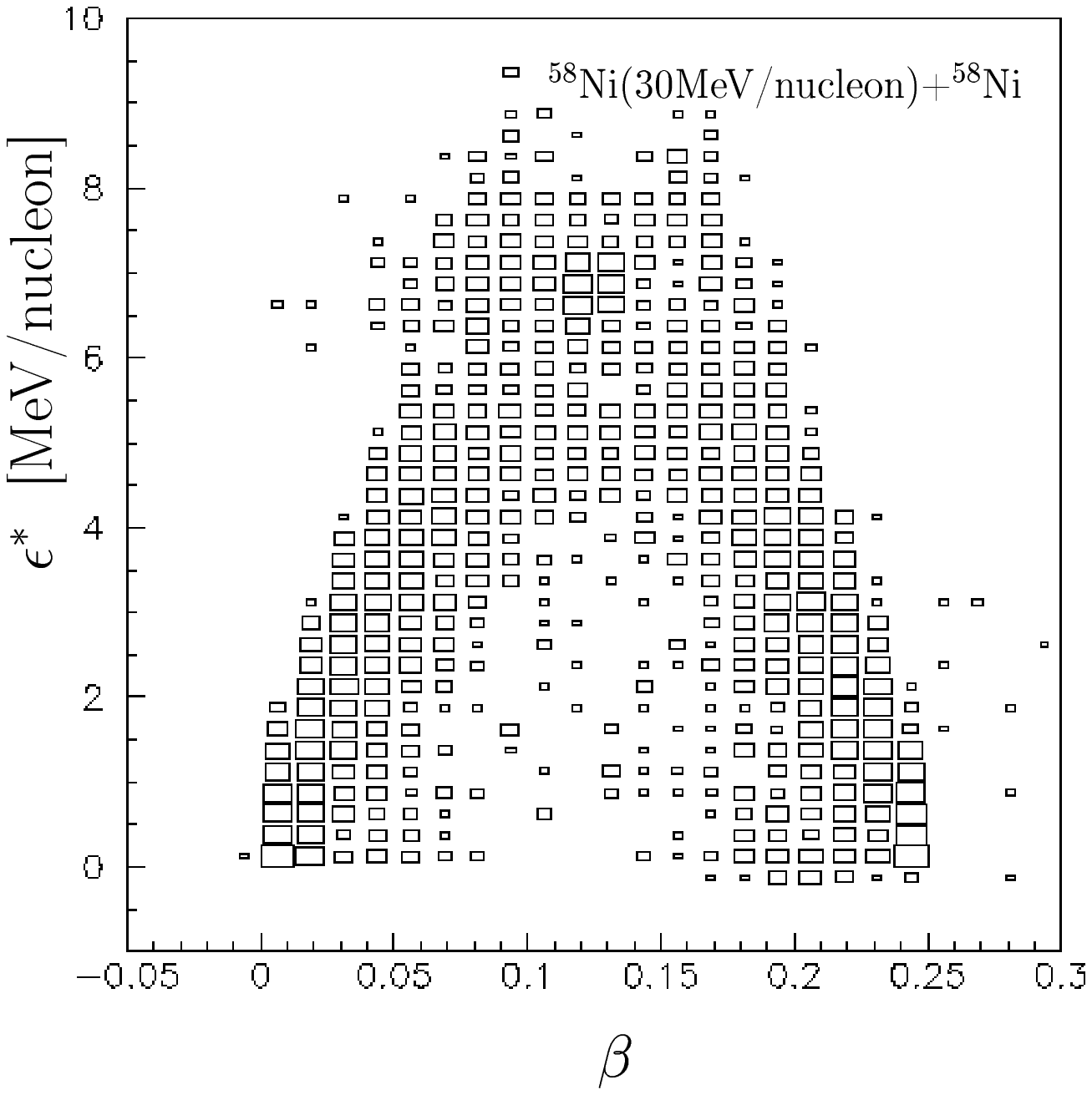}
\caption{ \footnotesize 
Calculated correlation of the excitation energy and velocity of the hot 
nuclei produced in the reaction $^{58}$Ni+$^{58}$Ni at 30 MeV/nucleon. 
}
\label{fgnini}
\end{figure}

The comparison to the broad range of experimental observables measured 
in various reactions in the Fermi energy domain appears 
to imply that the present approach describes correctly the processes 
leading to the production of excited projectile-like and target-like nuclei 
in the range between 20 and 50 MeV/nucleon. With increasing projectile 
energy, the production of three-body events and the compression phenomena 
start to play an important role. As a primary candidates for three-body 
events can be considered hot fragments where the relative motion of the 
participant zone leads to values of intrinsic angular momenta 
above the critical angular momentum for fusion. The compression 
phenomena can primarily take place in the events where the fragment 
with the mass close to the compound nucleus is produced 
with low intrinsic angular momentum. 

In conclusion, the approach presented in the present work 
appears to be a suitable tool to describe the production of hot nuclei 
which later undergo multifragmentation. Furthermore, production of 
rare beams in the Fermi energy domain can be addressed using this 
approach ( for instance, experimental studies are underway 
at the Cyclotron Institute of Texas A\&M University \cite{FrgExp} ). 

The author would like to thank to G.A. Souliotis and S.J. Yennello 
for support and fruitful and stimulating discussions and to L. Tassan-Got 
for the use of his DIT code. This work was supported in part by the NSF 
through grant No. PHY-9457376, 
the Robert A. Welch Foundation through grant No. A-1266, and 
the Department of Energy through grant No. DE-FG03-93ER40773. 
M.V. was also supported through grant VEGA-2/5121/98.


\begin{thebibliography}{99}

\bibitem{ExpRev}
C. Gregoire and B. Tamain: Ann. Phys. Fr. {\bf 11}, 323 (1986). 
\bibitem{LGPT}
J. Pochodzalla: Prog. Part. Nucl. Phys. {\bf 39}, 443 (1997). 
\bibitem{TrRev}
H. Feldmeier and J. Schnack: Rev. Mod. Phys. {\bf 72}, 655 (2000).
\bibitem{Griff}
J.J. Griffin: Phys. Rev. Lett. {\bf 17}, 478 (1966).  
\bibitem{HMB}
G. Harp, J. Miller, and B.J. Berne: Phys. Rev. {\bf 165}, 1166 (1968). 
\bibitem{RanVan}
J. Randrup and R. Vandenbosch: Nucl. Phys. A {\bf 474}, 219 (1987). 
\bibitem{VeZPA}  
M. Veselsky {\it et al.}:
Z. Phys. A {\bf 356}, 403 (1997).
\bibitem{Boh} 
M. B\"ohning: Nucl. Phys. A {\bf 152}, 529 (1970).
\bibitem{FuMo} 
H. Fuchs and K. M\"ohring: Rep. Prog. Phys. {\bf 57}, 231 (1994).
\bibitem{Park} 
D.J. Parker {\it et al.}:
Phys. Rev. C {\bf 44}, 1528 (1991).
\bibitem{TaGo}
L. Tassan-Got and C. St\'{e}fan, Nucl. Phys A {\bf 524}, 121 (1991).
\bibitem{VePRC}
M. Veselsky {\it et al.}: accepted to Phys. Rev. C, {\bf nucl-ex/0002007}.
\bibitem{SlRes}  
F.P. He\ss berger {\it et al.}: 
Z. Phys. A {\bf 348}, 301 (1994).
\bibitem{Si-Wil}  
J. Wilczynski {\it et al.}:
Nucl. Phys. A {\bf 373}, 109 (1982).
\bibitem{Tubbs}  
L.E. Tubbs {\it et al.}:
Phys. Rev. C {\bf 32}, 214 (1985).
\bibitem{Wilcz}
J. Wilczynski: Nucl. Phys. A {\bf 216}, 386 (1973). 
\bibitem{Gosset}
J. Gosset {\it et al.}:
Phys. Rev. C {\bf 16}, 629 (1977). 
\bibitem{Fried}
W.A. Friedman: Phys. Rev. C {\bf 27}, 569 (1983).
\bibitem{Gaim}
J.-J. Gaimard and K.-H. Schmidt: Nucl. Phys. A {\bf 531}, 709 (1991). 
\bibitem{Matsu} 
N. Matsuoka {\it et al.}:
Nucl. Phys. A {\bf 311}, 173 (1978).
\bibitem{Agni}
D.K. Agnihotri {\it et al.}: 
Adv. Nucl. Dyn. 1997, 3, 67. 
\bibitem{Holub}  
E. Holub {\it et al.}:
Phys. Rev. C {\bf 28}, 252 (1983).
\bibitem{Casini}
G. Casini {\it et al.}:
{\bf \hbox{nucl-ex/0010001}}.
\bibitem{Skul}
W. Skulski {\it et al.}:
Phys. Rev. C {\bf 53}, R2594 (1996). 
\bibitem{Sou}
G.A. Souliotis {\it et al.}: in preparation, see also 
G.A. Souliotis {\it et al.}: Phys. Rev. {\bf C 57}, 3129 (1998). 
\bibitem{GEM}
R. J. Charity {\it et al.}: Nucl. Phys. A {\bf 483}, 371 (1988).
\bibitem{Ramak}
E. Ramakrishnan {\it et al.}: Phys. Rev. C {\bf 57}, 1803 (1998). 
\bibitem{isodist} 
M. Veselsky {\it et al.}: Phys. Rev. C {\bf 62}, 41605 (2000).
\bibitem{mutemp} 
M. Veselsky {\it et al.}: submitted to Phys. Lett. B, 
{\bf \hbox{nucl-ex/0003004}}.
\bibitem{FrgExp}
G.A. Souliotis {\it et al.}: internal communication of the Cyclotron Institute. 




\end{thebibliography}
\end{document}